\documentclass[prd,twocolumn,aps,noshowpacs,nofootinbib,amsmath,amssymb,floatfix,superscriptaddress]{revtex4}
\usepackage{graphicx}
\usepackage{epstopdf}
\usepackage{bm}
\usepackage{amsmath}
\usepackage{amsfonts}
\usepackage{amssymb}
\usepackage{color}
\usepackage{multirow}
\usepackage{footnote}

\usepackage[colorlinks, citecolor=blue,anchorcolor=red,menucolor=red,linkcolor=red,filecolor=red,runcolor=red,urlcolor=blue,frenchlinks=red]{hyperref}
\begin{document}
\title{ Beauty-charm Meson Family with Coupled Channel Effects and Their Strong Decays }
\author{Wei Hao}
\affiliation{School of Physics, Nankai University, Tianjin 300071, China}

\author{Ruilin Zhu}
\email{rlzhu@njnu.edu.cn}
\affiliation{
Department of Physics and Institute of Theoretical Physics, Nanjing Normal University, Nanjing, Jiangsu 210023, China}
\affiliation{
Peng Huanwu Innovation Research Center, Institute of Theoretical Physics,
Chinese Academy of Sciences, Beijing 100190, China}

\begin{abstract}
We systematically study  the mass spectra and their two-body hadronic decays
of the beauty-charm meson family considering the coupled channel effects.
 Our results can good explain the observed $B_c$ meson spectrum and the prediction of the mass spectrum for unobserved beauty-charm mesons can be tested in future experiments. For the coupled channel components, we predicted the $1S$ state in beauty-charm meson family is about $4\%$, while the $2S$, $1P$, $2P$, $1D$, and $2D$ states are about $14\%$, $10\%$, $33\%$, and $17\%$ respectively. For the $3S$, $2P$ and $2D$ states, the  strong decay is allowed, The two-body hadronic decay widths of the $3^1S_0$, $3^3S_1$, $2^3P_2$ states are about 110 MeV, 69 MeV, and 3 MeV, respectively. While the two-body decay widths of the $2^3D_1$, $2D$, $2D^\prime$, and $2^3D_2$ states are 60 MeV, 149 MeV, 65 MeV, and 72 MeV, respectively.

\end{abstract}

\maketitle

\section{Introduction}

The understanding of hadron structures and their transitions at Fermi scale is fundamental issue from both the theoretical and experimental aspects in Particle Physics. For the conventional meson spectrum composed of a quark and an antiquark, the beauty-charm meson family is relatively incomplete.
Up to now, only three  beauty-charm mesons have been observed in experiments, i.e, the $B_c(1S)$, the $B_c(2S)$ and $B^*_c(2S)$.

The ground state $B_c(1^1S_0)$  of  beauty-charm family
was first discovered in 1998 by CDF at Fermilab~\cite{CDF:1998ihx}. The latest average mass for this state is $6274.47\pm0.27\pm0.17$ MeV~\cite{ParticleDataGroup:2022pth}.
One of the radial excited beauty-charm states was first discovered in $B_c^+ \pi^+ \pi^-$ invariant mass spectrum with subprocess $B_c^+\to J/\psi\pi^+$ using the sample corresponding to 4.9 fb$^{-1}$  of 7 TeV and 19.2 fb$^{-1}$ of
8 TeV $pp$ collision data collected by the ATLAS experiment at the LHC in 2014~\cite{ATLAS:2014lga}.
After that, two excited beauty-charm states $B_c(2^1S_0)$ and $B^*_c(2^3S_1)$ instead of one peak are discovered in  $B_c^+ \pi^+ \pi^-$ invariant mass spectrum from  both the CMS and LHCb experiments~\cite{CMS:2019uhm,LHCb:2019bem}. The combined average mass for the $B_c(2^1S_0)$ is determined as $6871.2\pm0.1$ MeV~\cite{ParticleDataGroup:2022pth}. For the vector excited state, $B^*_c(2^3S_1)$ first decays via hadronic transition $B^{*+}_c(2^3S_1) \to B^{*+}_c(1^3S_1) \pi^+ \pi^-$,  and then the vector $B^{*+}_c(1^3S_1)$ decays via electromagnetic transition $B^{*+}_c(1^3S_1)\to B^{+}_c(1^1S_0)+\gamma$. However, the radiated photon is soft with energy around 60MeV, which is not reconstructed in both the CMS and LHCb experiments. Thus the value of the mass of $B^{*+}_c(2^3S_1)$ relies on the precise information of $B^{*+}_c(1^3S_1)$. On the other hand, other beauty-charm meson states including orbitally excited states are
not observed at experiments yet.

In theoretical aspects, the mass spectra of beauty-charm mesons have been studied by many groups. For example, the quark potential models~\cite{Akbar:2019kbi,Li:2019tbn,Asghar:2019qjl,Akbar:2018hiw,Monteiro:2016rzi,Li:2022bre,Li:2023wgq,Gao:2024yvz,Godfrey:1985xj,Chang:2019wpt}, QCD sum rule~\cite{Dominguez:1993rg,Gershtein:1994dxw,Bagan:1994dy,Chen:2013eha,Wang:2012kw},  the heavy quark effective theory~\cite{Zeng:1994vj}, and the Dyson-Schwinger equation approach of QCD~\cite{Chang:2019eob,Chen:2020ecu}. Besides, the properties of the low-lying $B_c$ mesons are also investigated in the lattice QCD based on the first principles~\cite{Davies:1996gi,deDivitiis:2003iy,Allison:2004hy}.

The quark potential models in Refs.~\cite{Akbar:2019kbi,Li:2019tbn,Asghar:2019qjl,Akbar:2018hiw,Monteiro:2016rzi,Li:2022bre,Li:2023wgq,Godfrey:1985xj} are usually called quenched quark models. In quenched quark models, the convention mesons are constituted by a quark and an untiquark, and the mass spectrum comes from the interactions between constituent quarks. Therein the Godfrey-Isgur relativistic quark model~\cite{Godfrey:1985xj} is usually thought to provide a good description for most of the meson spectra. However, the quenched quark models sometimes explain poorly for higher excited states beyond the two-body threshold because they miss the generation of the light quark-antiquark pairs which enlarge the Fock space of the initial state~\cite{Lu:2016mbb}. These multiquark components will change the
Hamiltonian of the quark potential models, and then lead to mass shift and mixing among states with the same quantum numbers. If the initial state is above two-body threshold, the open channel strong decay will be allowed~\cite{Lu:2016mbb}. In other words, the unquenched quark model includes virtual hadronic loops. The hadronic loop has turned out to be highly nontrivial and  can give rise to mass shifts to the bare hadron states and contribute continuum components to the physical hadron states~\cite{Liu:2011yp}.

The coupled-channel model as one of the unquenched quark model, which is usually neglected, will manifest as a coupling to meson-meson (meson-baryon) channels and lead to mass shifts. The effects of quark-antiquark pairs are introduced explicitly into the constituent quark model via a QCD-inspired $^3P_0$ pair-creation mechanism. The approach is based on a constituent quark model to which the quark-antiquark pairs with vacuum quantum numbers are added as a perturbation. The pair-creation mechanism is inserted at the quark level and the one-loop diagrams are calculated by summing over the possible intermediate states~\cite{Ferretti:2012zz}.
It has been shown that the coupled-channel effects play an important role for describing the mesons spectra, such as charmonium~\cite{Kalashnikova:2005ui,Li:2009ad,Ferretti:2013faa,Deng:2023mza}, bottomonium~\cite{Liu:2011yp,Ferretti:2012zz,Ferretti:2013vua,Lu:2016mbb}, and charmed-strange mesons~\cite{vanBeveren:2003jv,vanBeveren:2003kd,Coito:2011qn,Hwang:2004cd,Simonov:2004ar,Lee:2004gt,Guo:2007up,Zhou:2011sp,Badalian:2007yr,Dai:2006uz,Hao:2022vwt}. In this paper, we will use this kind of unquenched quark model to study the $B_c$ mesons. we will investigate the mass spectrum of the beauty-charm mesons within the nonrelativistic  quark model by taking into account the mass shifts from the coupled-channel effect.

The paper is arranged as follows.
The theoretical formalism in coupled channel framework is given in  Section~\ref{sec:formalism}, where the nonrelativistic quenched  quark model and ${}^3P_0$ model are introduced.
The beauty-charm meson spectrum including  coupled channel  effects, the molecule and two quark components, and the two body hadronic decay widths are given in Section~\ref{sec:results}.
In the end, we give the summary  in Section~\ref{sec:summary}.

\section{Theoretical Formalism}\label{sec:formalism}

\subsection{Quenched quark model}
First we introduce a nonrelativistic quark model to denote the quenched quark interactions, which can be described in the Hamiltonian $H_Q$.
  \begin{eqnarray}
    H_Q& =& m_{b} + m_{\bar{c}} + \frac{\nabla^2}{2m_r}-C_F\frac{{\alpha}_s}{r}+br+C_{b\bar{c}}\nonumber\\  && + \frac{32{\alpha}_s{\sigma}^3 e^{-{\sigma}^2r^2}}{9\sqrt{\pi}m_bm_{\bar{c}}} {\boldsymbol{S}}_{b} \cdot {\boldsymbol{S}}_{\bar{c}}+H_{SL},\label{eqHA}
\end{eqnarray}
where the reduced quark mass $m_r$ satisfies $m_r=m_b m_{\bar{c}}/(m_b+m_{\bar{c}})$. Therein the ${\boldsymbol{S}}_{i}$ denotes the heavy quark spin
operator. The linear confining assumption is employed and the parameters
 $m_{b}$, $m_{\bar{c}}$, $C_{b\bar{c}}$, $\sigma$, $b$, and $\alpha_s$ in the quenched quark model Hamiltonian will be refitted from the knowledge
 of the existing hadrons.

The left spin and orbital related term $H_{SL}$ has the expression
    \begin{eqnarray}
      H_{SL} &=& \left(\frac{\boldsymbol{S}_{b}}{2m_b^2}+\frac{{\boldsymbol{S}}_{\bar{c}}}{2m_{\bar{c}}^2}\right) \cdot \boldsymbol{L}\left(\frac{1}{r}\frac{dV_c}{dr}+\frac{2}{r}\frac{dV_1}{dr}\right)\nonumber\\
      &&+\frac{{\boldsymbol{S}}_+ \cdot \boldsymbol{L}}{m_bm_{\bar{c}}}\left(\frac{1}{r} \frac{dV_2}{r}\right) \nonumber\\
      && +\frac{3{\boldsymbol{S}}_{b} \cdot \hat{\boldsymbol{r}}{\boldsymbol{S}}_{\bar{c}} \cdot \hat{\boldsymbol{r}}-{\boldsymbol{S}}_{b} \cdot {\boldsymbol{S}}_{\bar{c}}}{3m_b m_{\bar{c}}}V_3\nonumber\\
      && +\left[\left(\frac{{\boldsymbol{S}}_{b}}{m_b^2}-\frac{{\boldsymbol{S}}_{\bar{c}}}{m_{\bar{c}}^2}\right) \cdot \boldsymbol{L}+\frac{{\boldsymbol{S}}_-}{m_b m_{\bar{c}}} \cdot \boldsymbol{L}\right]  V_4,
\end{eqnarray}
where $\boldsymbol{L}$ is the orbital angular momentum between beauty charm quarks. $\boldsymbol{S}_{\pm}={\boldsymbol{S}}_b\pm{\boldsymbol{S}}_{\bar{c}}$.  The expressions for each potential are
\begin{eqnarray}
  V_c &=& -C_F\frac{{\alpha}_s}{r}+br,\nonumber \\
  V_1 &=& -br-\frac{2}{9\pi}\frac{{\alpha}_s^2}{r}[9{\rm ln}(\sqrt{m_b m_{\bar{c}}}r)+9{\gamma}_E-4],\nonumber\\
  V_2 &=& -C_F\frac{{\alpha}_s}{r}-\frac{1}{9\pi}\frac{{\alpha}_s^2}{r}[-18{\rm ln}(\sqrt{m_b m_{\bar{c}}}r)+54{\rm ln}(\mu r)\nonumber\\
  &&+36{\gamma}_E+29],\nonumber\\
  V_3 &=& -\frac{4{\alpha}_s}{r^3}-\frac{1}{3\pi}\frac{{\alpha}_s^2}{r^3}[-36{\rm ln}(\sqrt{m_b m_{\bar{c}}}r)+54{\rm ln}(\mu r)\nonumber\\
  &&+18{\gamma}_E+31],\nonumber\\
  V_4 &=& \frac{1}{\pi}\frac{{\alpha}_s^2}{r^3}{\rm ln}\left(\frac{m_{\bar{c}}}{m_b}\right),
\end{eqnarray}
where $\gamma_E$ denots Euler constant. The SU(3) color factors are $C_F=4/3$ and $C_A=3$.
The renormalization scale $\mu=1$~GeV is adopted as in Refs.~\cite{Lakhina:2006fy,Lu:2016bbk,Li:2010vx}.

The spin operator $\boldsymbol{S}_{-}={\boldsymbol{S}}_b-{\boldsymbol{S}}_{\bar{c}}$ will lead to the
mixing of the beauty-charm mesons with identical total angular momentum but with different total spins.
For example, there is mixing between $B_c(n{}^3L_L)$ and $B_c(n{}^1L_L)$ states, which can be described by
mixing matrix as follows by introducing a mixing angle $\theta_{nL}$~\cite{Godfrey:1985xj,Godfrey:1986wj}
\begin{equation}
\left(
\begin{array}{cr}
B_{cL}(nL)\\
B^\prime_{cL}(nL)
\end{array}
\right)
 =\left(
 \begin{array}{cr}
\cos \theta_{nL} & \sin \theta_{nL} \\
-\sin \theta_{nL} & \cos \theta_{nL}
\end{array}
\right)
\left(\begin{array}{cr}
B_c(n^1L_L)\\
B_c(n^3L_L)
\end{array}
\right),
\label{eqn:mix}
\end{equation}
where the physical observed states are denoted as $B_{cL}(nL)$ and $B_{cL}^\prime(nL)$.

\subsection{ Coupled Channel Framework}
The quenched quark potential model only considers the interaction between heavy quark pair, however, the hadronic loop interaction will also play a role
via the creation of the light quark pair in $b\bar{c}\to (b\bar{q})(q\bar{c})$. Furthermore, the hadronic loop interaction becomes much important for the higher excited beauty-charm mesons. The coupled channel framework provides a good description for the  hadronic loop interactions.

\begin{table}[h]
\caption{Parameters employed in the paper.}
\label{tab:para}
\begin{center}
\begin{tabular}{ccc}
\hline
\hline
Parameters  &    Fitted values    \\      
\hline
$m_n$      & $0.45$ GeV     \\ 
$m_s$      & $0.55$ GeV     \\ 
$m_c$      & $1.43$ GeV     \\
$m_b$      & $4.5$ GeV      \\
$\alpha_s$ & $0.51881$        \\
$b$        & $0.16164$ GeV$^2$ \\
$\sigma$   & $1.3381$ GeV     \\
$C_{bc}$   & $0.449$ GeV    \\ 
$\gamma_0$ & $0.4$          \\
\hline
\hline
\end{tabular}
\end{center}
\end{table}

\begin{table}[htpb]
\caption{\label{tab:spectrum} The beauty charm meson family spectrum (in MeV).
The third column denotes the naive mass in quenched quark model; the fourth column denotes the mass shift from coupled channels effects; the fifth column denotes the final results for the beauty charm meson family spectrum; the last column is the latest experimental data~\cite{ParticleDataGroup:2022pth}. The mixing angle of the $1P$, $2P$, $1D$ and $2D$ are $-35.0^\circ$, $-36.4^\circ$, $-42.8^\circ$ and $-44.1^\circ$, respectively.
}
\footnotesize
\begin{tabular}{ccccccc}
\hline\hline
  $n^{2S+1}L_J$  & State       &$M_0$  &$\Delta M$  &$M$    & PDG~\cite{ParticleDataGroup:2022pth}  \\\hline
  $1^1S_0$     & $B_c^\pm$       &$6332$  &$-60$  &$6272$  &$6274.47\pm0.32$    \\
  $1^3S_1$     & $-$           &$6399$  &$-64$  &$6335$  &$-$    \\
  $2^1S_0$     & $B^\pm_c(2S)$ &$6983$  &$-108$ &$6875$  &$6871.2\pm1.0$          \\
  $2^3S_1$     & $-$           &$7007$  &$-110$ &$6897$  &$-$    \\
  $3^1S_0$     & $-$           &$7388$  &$-137$ &$7250$  &$-$          \\
  $3^3S_1$     & $-$           &$7404$  &$-124$ &$7280$  &$-$    \\
  $1^3P_0$     & $-$           &$6798$  &$-91$  &$6707$  &$-$      \\
  $1P$         & $-$           &$6846$  &$-96$  &$6751$  &$-$       \\
  $1P^\prime$  & $-$           &$6881$  &$-95$  &$6786$  &$-$     \\
  $1^3P_2$     & $-$           &$6901$  &$-99$  &$6802$  &$-$     \\
  $2^3P_0$     & $-$           &$7220$  &$-122$ &$7099$  &$-$ \\
  $2P$         & $-$           &$7258$  &$-131$ &$7127$  &$-$  \\
  $2P^\prime$  & $-$           &$7294$  &$-129$ &$7165$  &$-$   \\
  $2^3P_2$     & $-$           &$7310$  &$-138$ &$7172$  &$-$ \\
  $1^3D_1$     & $-$           &$7142$  &$-117$ &$7025$  &$-$    \\
  $1D$         & $-$           &$7148$  &$-116$ &$7033$  &$-$    \\
  $1D^\prime$  & $-$           &$7157$  &$-116$ &$7041$  &$-$    \\
  $1^3D_3$     & $-$           &$7157$  &$-115$ &$7042$  &$-$       \\
  $2^3D_1$     & $-$           &$7489$  &$-129$ &$7360$  &$-$    \\
  $2D$         & $-$           &$7499$  &$-146$ &$7354$  &$-$    \\
  $2D^\prime$  & $-$           &$7503$  &$-127$ &$7376$  &$-$    \\
  $2^3D_3$     & $-$           &$7508$  &$-170$ &$7337$  &$-$       \\
  \hline\hline
\end{tabular}
\end{table}

In this framework, the beauty-charm meson state can be written as
\begin{align}
    |\psi\rangle = \left(\begin{array}{cc}
                           c_0 |\psi_0\rangle  \\
                           \sum_{BD} \int d^3p\, c_{BD}(p) |BD;p\rangle
                         \end{array}\right),
\end{align}
where $c_0$ is the $b\bar{c}$ bare state probability amplitude, while
$c_{BD}(p)$ is the beauty meson $B$ and charm meson $D$ molecular component probability amplitude 
with relative momentum $p$. 
To normalize the state, we have the condition $|c_0|^2+\sum_{BD} \int d^3p\, |c_{BD}(p)|^2=1$.

The total Hamiltonian in coupled channel framework is 
\begin{align}
    H = \left(\begin{array}{cc}
                          H_Q & H_I  \\
                          H_I & H_{BD}
                         \end{array}\right),
\end{align}  
where $H_{BD}$ is the Hamiltonian for beauty meson $B$ and charm meson $D$ system as
\begin{align}
    H_{BD} = E_{BD} =\sqrt{m_B^2 +p^2} + \sqrt{m_D^2 +p^2}.
\end{align}
$H_I$ leads to the coupling between $b\bar{c}$ bare state and BD molecule component.
In the following, the $^3P_0$ model where the generated light quark pairs have  identical quantum numbers $J^{PC} = 0^{++}$ with vacuum is employed to analyze 
the   mixing of $b\bar{c}$ bare state and BD molecule component~\cite{Micu:1968mk, LeYaouanc:1972vsx, LeYaouanc:1973ldf}.

Then we solve the spectrum eigen equation
\begin{align}
    H  |\psi\rangle= M  |\psi\rangle,
\end{align}
where $M$ is the final mass for the beauty-charm mesons in coupled channel framework. Practically the eigenvalue $M$ can be rewritten as~\cite{Kalashnikova:2005ui},
\begin{align}
\label{m}
M &= M_Q + \Delta M, \\
\Delta M &= \sum_{BD} \int_0^{\infty} p^2 dp \frac{\left|\left\langle BD;p \right| T^\dagger \left| \psi_0 \right\rangle \right|^2}{M - E_{BD} + i\epsilon}
\end{align}
where $M_Q$ is the eigenvalue for the quenched Hamiltonian $H_Q$ while $\Delta M$ is the mass shift from the coupled channel effect.
The operator $T^\dag$ in the $^3P_0$ model can be written as~\cite{Ferretti:2013faa,Ferretti:2012zz,Ferretti:2013vua}
\begin{equation}
	\label{eqn:Tdag}
	\begin{array}{rcl}
	T^{\dagger} &=& -3 \, \gamma_0^{eff} \, \int d \vec{p}_1 \, d \vec{p}_2 \,
	\delta(\vec{p}_1 + \vec{p}_2) \, C_{12} \, F_{12} \,
	{e}^{-r_q^2 (\vec{p}_1 - \vec{p}_2)^2/6 }\,  \\
	& & \left[ \chi_{12} \, \times \, {\cal Y}_{1}(\vec{p}_1 - \vec{p}_2) \right]^{(0)}_0 \,
	b_1^{\dagger}(\vec{p}_1) \, d_2^{\dagger}(\vec{p}_2) ~,
	\end{array}
\end{equation}
where the operators $ b_1^{\dagger}(\vec{p}_1)$ and $d_2^{\dagger}(\vec{p}_2)$ creates a light quark pair.
The light quark pair creation strength is denoted as $\gamma_0^{eff}=\frac{m_u}{m_i}\gamma_0$ with $\gamma_0=0.4$
and $i=u, d, s$~\cite{Li:2019tbn}.
The color, flavor and spin wave  functions for the light quark pair are 
$C_{34}$, $F_{34}$ and $\chi_{34}$, respectively. The parameter $r_q$ in the Gaussian factor to describe the quark pair creation is in the range $0.25$ to $0.35$ fm~\cite{Silvestre-Brac:1991qqx,Geiger:1991ab,Geiger:1991qe,Geiger:1996re}.
We will use the value $r_q = 0.3$ fm in the following calculation.

To weight the importance of coupled channel effects, it is useful to investigate the 
probabilities of the $b\bar{c}$ bare component and BD molecule component in the physical state. 
 The probability of quenched $b\bar{c}$ bare component is given as
\begin{eqnarray}
\label{eqn:pqqbar}
	P_{b\bar{c}} &\equiv& |c_0|^2 \nonumber\\
&= &\left(1+\sum_{BD} \int_0^{\infty} p^2 dp \frac{\left|\left\langle BD;p \ell J \right| T^\dagger \left| \psi_0 \right\rangle \right|^2}{(M - E_{BD})^2}\right)^{-1}.\nonumber\\
\end{eqnarray}
Then the probability of BD molecule component naturally expressed as $P_{\mathrm{molecule}}=\sum_{BD}P_{BD}= 1- P_{b\bar{c}}$.

For higher excited beauty-charm mesons above the $BD$ threshold, they directly two-body decay into beauty meson and charm meson.
The strong decay width is related to the imaginary part in $\Delta M$ and can be written as
\begin{equation}
\label{eqn:decay}
    \Gamma_{BD} = 2 \pi p_0 \frac{E_B(p_0) E_D(p_0)}{M} \left| \left\langle BD;p_0\right| T^\dagger \left| \psi_0 \right\rangle \right|^2
\end{equation}

\begin{table*}
\caption{\label{tab:shift} Mass shift $\Delta M$ (in MeV) for beauty charm mesons from different channels.}
\begin{tabular}{cccccccccccc}
\hline
\hline
State                &$BD$  &$BD^*$ &$B^*D$ &$B^*D^*$ &$B_sD_s$ &$B_sD_s^*$ &$B_s^*D_s$   &$B_s^*D_s^*$  &Total \\
\hline
$1^1S_0$             &$0$    &$-11$    &$-12$   &$-22$  &$0$   &$-4$    &$-4$    &$-8$      &$-60$  \\
$1^3S_1$             &$-4$   &$-8$     &$-9$    &$-28$  &$-1$  &$-3$    &$-3$    &$-9$      &$-64$  \\
$2^1S_0$             &$0$  &$-20$    &$-24$   &$-41$  &$0$   &$-5$    &$-6$    &$-11$    &$-108$  \\
$2^3S_1$             &$-8$    &$-14$    &$-16$   &$-49$  &$-2$  &$-4$   &$-4$     &$-13$     &$-110$ \\
$3^1S_0$             &$0$  &$-34$    &$-25$   &$-57$  &$0$   &$-5$    &$-6$    &$-10$    &$-136$  \\
$3^3S_1$             &$3$    &$-30$    &$2$   &$-75$  &$-2$  &$-3$   &$-4$     &$-12$     &$-122$ \\
$1^3P_0$             &$-12$  &$0$      &$0$     &$-58$  &$-3$  &$0$   &$0$      &$18$      &$-91$  \\
$1P$                 &$0$    &$-18$    &$-20$   &$-36$  &$0$   &$-5$    &$-6$    &$-11$     &$-95$  \\
$1P^\prime$          &$0$    &$-14$    &$-16$   &$-43$  &$0$   &$-4$    &$-5$    &$-13$    &$-95$  \\
$1^3P_2$             &$-10$  &$-13$    &$-15$   &$-39$  &$-3$  &$-4$   &$-4$    &$-11$     &$-99$  \\
$2^3P_0$             &$-33$   &$0$      &$0$     &$-68$  &$-4$  &$0$   &$0$      &$-17$      &$-122$  \\
$2P$                 &$0$    &$-25$    &$-35$    &$-48$  &$0$   &$-5$    &$-6$     &$-11$     &$-131$  \\
$2P^\prime$          &$0$    &$-23$    &$-34$   &$-51$   &$0$   &$-5$    &$-5$     &$-12$     &$-129$  \\
$2^3P_2$             &$-19$  &$-17$    &$-22$   &$-58$   &$-3$  &$-4$   &$-4$     &$-12$     &$-138$  \\
$1^3D_1$             &$-13$   &$-4$    &$-6$     &$-70$  &$-3$ &$-1$   &$-1$     &$-19$     &$-117$  \\
$1D$                 &$0$    &$-22$    &$-26$   &$-44$  &$0$   &$-6$    &$-6$    &$-12$     &$-116$  \\
$1D^\prime$          &$0$    &$-19$    &$-23$   &$-50$  &$0$   &$-5$    &$-5$    &$-14$     &$-116$  \\
$1^3D_3$             &$-13$  &$-16$    &$-18$   &$-44$  &$-3$  &$-4$   &$-5$    &$-11$     &$-115$ \\
$2^3D_1$             &$-6$   &$1$    &$1$     &$-99$  &$-6$ &$-1$   &$-2$     &$-16$     &$-129$  \\
$2D$                 &$0$    &$-26$    &$-18$   &$-80$  &$0$   &$-5$    &$-6$    &$-10$     &$-145$  \\
$2D^\prime$          &$0$    &$-11$    &$-12$   &$-76$  &$0$   &$-5$    &$-7$    &$-12$     &$-123$  \\
$2^3D_3$             &$-10$  &$-21$    &$-20$   &$-98$  &$-3$  &$-3$   &$-4$    &$-10$     &$-169$ \\
\hline
\hline
\end{tabular}
\end{table*}

\begin{table*}
\caption{\label{tab:Pro} The two quark and molecule probabilities (in $\%$) in the coupled channels framework.}
\begin{tabular}{cccccccccccc}
\hline
\hline
State                &$BD$  &$BD^*$ &$B^*D$ &$B^*D^*$ &$B_sD_s$ &$B_sD_s^*$ &$B_s^*D_s$   &$B_s^*D_s^*$  &$P_{molecule}$  &$P_{b\bar c}$ \\
\hline
$1^1S_0$             &$0$    &$0.7$    &$0.8$   &$1.4$  &$0$   &$0.2$    &$0.2$    &$0.4$      &$3.7$  &$96.3$  \\
$1^3S_1$             &$0.3$   &$0.5$     &$0.6$    &$1.8$  &$0.1$  &$0.1$    &$0.2$    &$0.5$      &$4.3$  &$95.7$ \\
$2^1S_0$             &$0$  &$2.8$    &$3.9$   &$5.1$  &$0$   &$0.5$    &$0.7$    &$1.1$    &$14.0$  &$86.0$ \\
$2^3S_1$             &$1.6$    &$1.9$    &$2.7$   &$6.2$  &$0.2$  &$0.4$   &$0.5$     &$1.3$     &$14.8$  &$85.2$\\
$1^3P_0$             &$1.8$  &$0$      &$0$     &$5.2$  &$0.4$  &$0$   &$0$      &$1.3$      &$8.6$   &$91.4$\\
$1^1P_1$             &$0$    &$1.8$    &$2.4$   &$3.5$  &$0$   &$0.4$    &$0.5$    &$0.9$     &$9.6$   &$90.4$\\
$1^3P_1$             &$0$    &$1.6$    &$2.1$   &$4.0$  &$0$   &$0.4$    &$0.4$    &$1.0$    &$9.5$   &$90.5$\\
$1^3P_2$             &$1.2$  &$1.3$    &$1.7$   &$4.1$  &$0.3$  &$0.3$   &$0.4$    &$1.0$     &$10.2$   &$89.8$\\
$2^3P_0$             &$22.6$   &$0$      &$0$     &$9.0$  &$0.7$  &$0$   &$0$      &$1.4$      &$33.7$   &$66.3$\\
$2^1P_1$             &$0$    &$5.3$    &$16.8$    &$8.4$  &$0$   &$0.6$    &$0.8$     &$1.1$     &$32.9$   &$67.1$\\
$2^3P_1$             &$0$    &$5.3$    &$18.3$   &$7.5$   &$0$   &$0.5$    &$0.7$     &$1.1$     &$33.4$   &$66.6$\\
$1^3D_1$             &$4.3$   &$0.9$    &$1.5$     &$8.1$  &$0.4$ &$0.1$   &$0.2$     &$1.7$     &$17.3$   &$82.7$\\
$1^1D_2$             &$0$    &$3.2$    &$4.7$   &$5.8$  &$0$   &$0.6$    &$0.7$    &$1.1$     &$16.2$    &$83.8$\\
$1^3D_2$             &$0$    &$3.1$    &$4.8$   &$6.2$  &$0$   &$0.5$    &$0.7$    &$1.2$     &$16.5$   &$83.5$\\
$1^3D_3$             &$2.2$  &$2.0$    &$2.6$   &$6.6$  &$0.4$  &$0.4$   &$0.5$    &$1.2$     &$15.9$    &$84.1$\\
\hline
\hline
\end{tabular}
\end{table*}

\section{Results and discussions}
\label{sec:results}

 Because only two $B_c$ mesons have experimental information, it is difficult to fit all parameters. So We adopt a strategy similar to that in Ref.~\cite{Li:2019tbn}. In our calculation, the parameters $\alpha_s$, $b$ and $\sigma$ are determined by fitting the mass spectrum of the $B_c$, $B_c^*$ and $B^{(*)}_c(2S)$, and the other mesons come from previous papers~\cite{Li:2010vx,Li:2019tbn,Lakhina:2006fy}.  Though there are no experimental data for the $B^*_c(1^3S_1)$ and $B^*_c(2^3S_1)$ masses, but their values can be estimate to be around $6334$~MeV and $6900$~MeV, respectively. Because the hyperfine mass splitting in bottomonium family is measured as $\Delta M_{b\bar{b}(1S)}=62.3\pm 3.2$ MeV and  $\Delta M_{b\bar{b}(2S)}=24\pm4$ MeV~\cite{ParticleDataGroup:2022pth}, the hyperfine mass splitting in beauty-charm meson family is believed to be smaller as $m_{B_c^*}-m_{B_c}\geq\Delta M_{b\bar{b}(1S)}=62.3\pm 3.2$ MeV and $m_{B_c^*(2S)}-m_{B_c(2S)}\geq\Delta M_{b\bar{b}(2S)}=24\pm4$ MeV since the  hyperfine mass splitting is inversely proportional to the heavy quark mass. The final refitted parameters are listed in Table~\ref{tab:para}.

\begin{figure}
    \centering
    \includegraphics[width=\columnwidth]{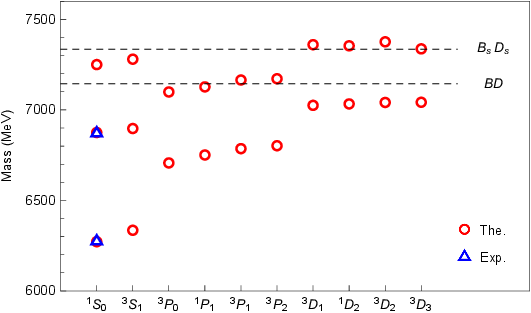}
    \caption{The beauty-charm meson family spectrum.  ``Exp.'' denotes the current experimental values from the latest PDG~\cite{ParticleDataGroup:2022pth} and our theoretical results are depicted as ``The.''. The dashed lines represent the threshold positions of $BD$ and $B_sD_s$, respectively}
    \label{fig:spectrum}
\end{figure}

With the parameters in Table~\ref{tab:para},  the mass spectrum and mass shifts of the $B_c$ mesons can  be estimated. The results are shown in Fig.~\ref{fig:spectrum}, with numbers listed in Table~\ref{tab:spectrum}.  The mixing angle of the $1P$, $2P$, $1D$ and $2D$ states can be also  calculated, which are  $-35.0^\circ$,$-36.4^\circ$,$-42.8^\circ$ and $-44.1^\circ$, respectively. The mixing angles are close to the  $B$ mesons $\theta_{1P}=34.6$, $\theta_{2P}=-36.1^\circ$, $\theta_{1D}=-39.6^\circ$, $\theta_{2D}=-39.7^\circ$ and $B_s$ mesons $\theta_{1P}=34.9$, $\theta_{2P}=-36.1^\circ$, $\theta_{1D}=-39.8^\circ$, $\theta_{2D}=-39.8^\circ$~\cite{Lu:2016bbk}.

We predicted that the masses of the $B_c(1^1S_0)$ and the $B_c(2^1S_0)$ are 6272 MeV and 6875 MeV, which are close to the experimental values $6274.47\pm0.32$ MeV and $6871.2\pm1.0$ MeV. Besides, we also show the theoretical results of the other states in beauty-charm meson family. We expect more experimental information can be found to support our results.

For states below $BD$ threshold, the probabilities of each coupled channel can be estimated. The probabilities are listed in Table \ref{tab:Pro}. Therein, all the states have coupled channel components. Especially when comparing with the $1S$, $2S$, $1P$, and the $1D$ states, the $2P$ states have larger non-$b\bar c$ components. For the $1S$-wave states, we predicted they have $96\%$ $b\bar c$ components, that means the coupled channel components is just $4\%$. For other states, we predicted the $b\bar c$ component probabilities for the $2S$, $1P$, $2P$, $1D$, and $2D$ states are about $86\%$, $90\%$, $67\%$, and $83\%$, respectively.

For states which have large masses, the strong decay channels will be open, the strong decay width are shown in Table~\ref{tab:width}. For the $1S$-wave, $2S$-wave, $1P$-wave, $2P$-wave and $1D$-wave states, their masses are below the $BD$ threshold and can not strong decay into $BD$ states. So we just discuss the strong decay of the $B_c(2^3P_2)$, $3S$-wave and the $2D$-wave $B_c$ mesons here.  For the $B_c(2^3P_2)$, it can just decay to $BD$ final states with decay width 3 MeV. For the $3S$-wave states, the $B_c(3^1S_0)$ can strong decay to $B^*D$ with predicted width 110 MeV, while the $B_c(3^3S_1)$ can strong decay to $BD$ and $B^*D$ with predicted width 10 MeV and 59 MeV, whose total decay width becomes 69 MeV. For the $2D$-wave states, the total decay widths of the $B_c(2^3D_1)$, $B_c(2D)$, $B_c(2D^\prime)$ and $2^3D_3$ states are $60$ MeV, $149$ MeV, $65$ MeV and $72$ MeV, respectively. The $B_c(2^3D_1)$ dominantly decay into $BD^*$ and $B^*D^*$ with predicted widths $28$ MeV and $22$ MeV. The $B_c(2D)$ can mainly decay into $B^*D$ and $BD^*$ with predicted widths $92$ MeV and $54$ MeV. And the $B_c(2D^\prime)$ can mainly decay into $BD^*$and $B^*D^*$ with predicted widths $18$ MeV and $44$ MeV. The mixing angle of the two states is $-44.1^\circ$. The $B_c(2^3D_3)$ have three major decay channels $BD$,  $B^*D$ and $B^*D^*$, with the decay width $27$ MeV, $30$ MeV and $11$ MeV, respectively. These difference in decays will be helpful to distinguish these excited beauty-charm meson states.
\begin{table}[thb]
\caption{\label{tab:width} Hadronic decay widths  (in MeV) of the beauty-charm mesons.}
\begin{tabular}{cccccccccccc}
\hline
\hline
State                &$BD$  &$BD^*$ &$B^*D$ &$B^*D^*$ &$B_sD_s$ &$B_sD_s^*$ &$B_s^*D^{(*)}_s$   &Total \\
\hline
$3^1S_0$             &$0$  &$0$    &$110$   &$0$   &$0$  &$0$   &$0$        &$110$  \\
$3^3S_1$             &$10$  &$0$    &$59$   &$0$   &$0$  &$0$   &$0$          &$69$  \\
$2^3P_2$             &$3$  &$0$    &$0$   &$0$   &$0$  &$0$   &$0$        &$3$  \\
$2^3D_1$             &$2$   &$28$    &$1$     &$22$  &$7$ &$0$   &$0$         &$60$  \\
$2D$                 &$0$    &$92$    &$54$   &$3$  &$0$   &$0$    &$0$        &$149$  \\
$2D^\prime$          &$0$    &$18$    &$4$   &$44$  &$0$   &$0$    &$0$        &$65$  \\
$2^3D_3$             &$27$  &$4$    &$30$   &$11$  &$0$  &$0$   &$0$        &$72$ \\
\hline
\hline
\end{tabular}
\end{table}

\section{Summary}
\label{sec:summary}

We calculated the mass spectrum and two-body hadronic decays for beauty-charm mesons based on the coupled channel framework. The coupled channel effects are calculated from the $^3P_0$ model. The wave functions in our calculations are obtained by solving the Hamiltonian of the potential model with Gaussian Expansion Method.

Our results support all of the beauty-charm states have coupled channel components. Each component is different for various states. Generally, the coupled channel effects are smaller for bound states than excited states. The $1S$ states are about $3 \sim 5 \%$, while the $1P$ states are about $8 \sim 11 \%$. The $1D$ states are about $15 \sim 18 \%$, and the $2S$ states are about $14 \sim 15 \%$. Four $2P$ states have larger couple channel components around $32 \sim 34 \%$.

For the strong decays of excited states above $BD$ threshold, the $B_c(3^1S_0)$ state mainly strong decays to $B^*D$ channel and the $B_c(3^3S_1)$ mainly strong decays to $BD$ and $B^*D$. While the $B_c(2^3P_2)$ mainly decays to $BD$. In addition, the $B_c(3S)$ and $B_c^*(3S)$ can be also detected in $B_c(3S)\to B_c(1S/2S)+\pi^++\pi^-$ and $B^*_c(3S)\to B_c(1S/2S)+\pi^++\pi^-+\gamma$ processes practically. For four $D$-wave states, the $B_c(2D)$ dominantly decays to $BD^*$ and $B^*D$ final states with total width 149 MeV. However, the $B_c(2^3D_1)$, $B_c(2D^\prime)$, and $B_c(2^3D_3)$ states have smaller decay widths around 60 $\sim$ 72 MeV.
For the electromagnetical and weak decays of beauty-charm mesons with polarization analysis, one can also refer to Refs.~\cite{Geng:2023ffc,Tao:2022qxa,Wang:2018duy,Zhu:2017lwi,Qiao:2012hp}.

Only a few states of beauty-charm mesons are observed in current experiments, the theoretical studies will be useful to reveal their nature, and promote future experimental findings.

\section{Acknowledgements}
 This work is supported by NSFC under grant No.~12322503, No.~12047503, and No.~12075124,
 and by Natural Science Foundation of Jiangsu under Grant No.~BK20211267.

\bibliography{cite}  

\begin{thebibliography}{61}
\expandafter\ifx\csname natexlab\endcsname\relax\def\natexlab#1{#1}\fi
\expandafter\ifx\csname bibnamefont\endcsname\relax
  \def\bibnamefont#1{#1}\fi
\expandafter\ifx\csname bibfnamefont\endcsname\relax
  \def\bibfnamefont#1{#1}\fi
\expandafter\ifx\csname citenamefont\endcsname\relax
  \def\citenamefont#1{#1}\fi
\expandafter\ifx\csname url\endcsname\relax
  \def\url#1{\texttt{#1}}\fi
\expandafter\ifx\csname urlprefix\endcsname\relax\def\urlprefix{URL }\fi
\providecommand{\bibinfo}[2]{#2}
\providecommand{\eprint}[2][]{\url{#2}}

\bibitem[{\citenamefont{Abe et~al.}(1998)}]{CDF:1998ihx}
\bibinfo{author}{\bibfnamefont{F.}~\bibnamefont{Abe}} \bibnamefont{et~al.}
  (\bibinfo{collaboration}{CDF}), \bibinfo{journal}{Phys. Rev. Lett.}
  \textbf{\bibinfo{volume}{81}}, \bibinfo{pages}{2432} (\bibinfo{year}{1998}),
  \eprint{hep-ex/9805034}.

\bibitem[{\citenamefont{Workman et~al.}(2022)}]{ParticleDataGroup:2022pth}
\bibinfo{author}{\bibfnamefont{R.~L.} \bibnamefont{Workman}}
  \bibnamefont{et~al.} (\bibinfo{collaboration}{Particle Data Group}),
  \bibinfo{journal}{PTEP} \textbf{\bibinfo{volume}{2022}},
  \bibinfo{pages}{083C01} (\bibinfo{year}{2022}).

\bibitem[{\citenamefont{Aad et~al.}(2014)}]{ATLAS:2014lga}
\bibinfo{author}{\bibfnamefont{G.}~\bibnamefont{Aad}} \bibnamefont{et~al.}
  (\bibinfo{collaboration}{ATLAS}), \bibinfo{journal}{Phys. Rev. Lett.}
  \textbf{\bibinfo{volume}{113}}, \bibinfo{pages}{212004}
  (\bibinfo{year}{2014}), \eprint{1407.1032}.

\bibitem[{\citenamefont{Sirunyan et~al.}(2019)}]{CMS:2019uhm}
\bibinfo{author}{\bibfnamefont{A.~M.} \bibnamefont{Sirunyan}}
  \bibnamefont{et~al.} (\bibinfo{collaboration}{CMS}), \bibinfo{journal}{Phys.
  Rev. Lett.} \textbf{\bibinfo{volume}{122}}, \bibinfo{pages}{132001}
  (\bibinfo{year}{2019}), \eprint{1902.00571}.

\bibitem[{\citenamefont{Aaij et~al.}(2019)}]{LHCb:2019bem}
\bibinfo{author}{\bibfnamefont{R.}~\bibnamefont{Aaij}} \bibnamefont{et~al.}
  (\bibinfo{collaboration}{LHCb}), \bibinfo{journal}{Phys. Rev. Lett.}
  \textbf{\bibinfo{volume}{122}}, \bibinfo{pages}{232001}
  (\bibinfo{year}{2019}), \eprint{1904.00081}.

\bibitem[{\citenamefont{Akbar}(2020)}]{Akbar:2019kbi}
\bibinfo{author}{\bibfnamefont{N.}~\bibnamefont{Akbar}},
  \bibinfo{journal}{Phys. Atom. Nucl.} \textbf{\bibinfo{volume}{83}},
  \bibinfo{pages}{634} (\bibinfo{year}{2020}), \eprint{1911.02078}.

\bibitem[{\citenamefont{Li et~al.}(2019)\citenamefont{Li, Liu, Lu, L\"u, Gui,
  and Zhong}}]{Li:2019tbn}
\bibinfo{author}{\bibfnamefont{Q.}~\bibnamefont{Li}},
  \bibinfo{author}{\bibfnamefont{M.-S.} \bibnamefont{Liu}},
  \bibinfo{author}{\bibfnamefont{L.-S.} \bibnamefont{Lu}},
  \bibinfo{author}{\bibfnamefont{Q.-F.} \bibnamefont{L\"u}},
  \bibinfo{author}{\bibfnamefont{L.-C.} \bibnamefont{Gui}}, \bibnamefont{and}
  \bibinfo{author}{\bibfnamefont{X.-H.} \bibnamefont{Zhong}},
  \bibinfo{journal}{Phys. Rev. D} \textbf{\bibinfo{volume}{99}},
  \bibinfo{pages}{096020} (\bibinfo{year}{2019}), \eprint{1903.11927}.

\bibitem[{\citenamefont{Asghar et~al.}(2019)\citenamefont{Asghar, Akram, Masud,
  and Sultan}}]{Asghar:2019qjl}
\bibinfo{author}{\bibfnamefont{I.}~\bibnamefont{Asghar}},
  \bibinfo{author}{\bibfnamefont{F.}~\bibnamefont{Akram}},
  \bibinfo{author}{\bibfnamefont{B.}~\bibnamefont{Masud}}, \bibnamefont{and}
  \bibinfo{author}{\bibfnamefont{M.~A.} \bibnamefont{Sultan}},
  \bibinfo{journal}{Phys. Rev. D} \textbf{\bibinfo{volume}{100}},
  \bibinfo{pages}{096002} (\bibinfo{year}{2019}), \eprint{1910.02680}.

\bibitem[{\citenamefont{Akbar et~al.}(2019)\citenamefont{Akbar, Akram, Masud,
  and Atif~Sultan}}]{Akbar:2018hiw}
\bibinfo{author}{\bibfnamefont{N.}~\bibnamefont{Akbar}},
  \bibinfo{author}{\bibfnamefont{F.}~\bibnamefont{Akram}},
  \bibinfo{author}{\bibfnamefont{B.}~\bibnamefont{Masud}}, \bibnamefont{and}
  \bibinfo{author}{\bibfnamefont{M.}~\bibnamefont{Atif~Sultan}},
  \bibinfo{journal}{Eur. Phys. J. A} \textbf{\bibinfo{volume}{55}},
  \bibinfo{pages}{82} (\bibinfo{year}{2019}), \eprint{1811.07552}.

\bibitem[{\citenamefont{Monteiro et~al.}(2017)\citenamefont{Monteiro, Bhat, and
  Vijaya~Kumar}}]{Monteiro:2016rzi}
\bibinfo{author}{\bibfnamefont{A.~P.} \bibnamefont{Monteiro}},
  \bibinfo{author}{\bibfnamefont{M.}~\bibnamefont{Bhat}}, \bibnamefont{and}
  \bibinfo{author}{\bibfnamefont{K.~B.} \bibnamefont{Vijaya~Kumar}},
  \bibinfo{journal}{Phys. Rev. D} \textbf{\bibinfo{volume}{95}},
  \bibinfo{pages}{054016} (\bibinfo{year}{2017}), \eprint{1608.05782}.

\bibitem[{\citenamefont{Li et~al.}(2023{\natexlab{a}})\citenamefont{Li, Tang,
  Fang, Wang, Pang, and Liu}}]{Li:2022bre}
\bibinfo{author}{\bibfnamefont{T.-y.} \bibnamefont{Li}},
  \bibinfo{author}{\bibfnamefont{L.}~\bibnamefont{Tang}},
  \bibinfo{author}{\bibfnamefont{Z.-y.} \bibnamefont{Fang}},
  \bibinfo{author}{\bibfnamefont{C.-h.} \bibnamefont{Wang}},
  \bibinfo{author}{\bibfnamefont{C.-q.} \bibnamefont{Pang}}, \bibnamefont{and}
  \bibinfo{author}{\bibfnamefont{X.}~\bibnamefont{Liu}},
  \bibinfo{journal}{Phys. Rev. D} \textbf{\bibinfo{volume}{108}},
  \bibinfo{pages}{034019} (\bibinfo{year}{2023}{\natexlab{a}}),
  \eprint{2204.14258}.

\bibitem[{\citenamefont{Li et~al.}(2023{\natexlab{b}})\citenamefont{Li, Li,
  Wang, and Liu}}]{Li:2023wgq}
\bibinfo{author}{\bibfnamefont{X.-J.} \bibnamefont{Li}},
  \bibinfo{author}{\bibfnamefont{Y.-S.} \bibnamefont{Li}},
  \bibinfo{author}{\bibfnamefont{F.-L.} \bibnamefont{Wang}}, \bibnamefont{and}
  \bibinfo{author}{\bibfnamefont{X.}~\bibnamefont{Liu}}, \bibinfo{journal}{Eur.
  Phys. J. C} \textbf{\bibinfo{volume}{83}}, \bibinfo{pages}{1080}
  (\bibinfo{year}{2023}{\natexlab{b}}), \eprint{2308.07206}.

\bibitem[{\citenamefont{Gao et~al.}(2024)\citenamefont{Gao, Fan, Chen, and
  Pang}}]{Gao:2024yvz}
\bibinfo{author}{\bibfnamefont{Z.-b.} \bibnamefont{Gao}},
  \bibinfo{author}{\bibfnamefont{Y.-y.} \bibnamefont{Fan}},
  \bibinfo{author}{\bibfnamefont{H.}~\bibnamefont{Chen}}, \bibnamefont{and}
  \bibinfo{author}{\bibfnamefont{C.-q.} \bibnamefont{Pang}}
  (\bibinfo{year}{2024}), \eprint{2402.10629}.

\bibitem[{\citenamefont{Godfrey and Isgur}(1985)}]{Godfrey:1985xj}
\bibinfo{author}{\bibfnamefont{S.}~\bibnamefont{Godfrey}} \bibnamefont{and}
  \bibinfo{author}{\bibfnamefont{N.}~\bibnamefont{Isgur}},
  \bibinfo{journal}{Phys. Rev. D} \textbf{\bibinfo{volume}{32}},
  \bibinfo{pages}{189} (\bibinfo{year}{1985}).

\bibitem[{\citenamefont{Chang et~al.}(2021)\citenamefont{Chang, Chen, Li, Liu,
  and Raya}}]{Chang:2019wpt}
\bibinfo{author}{\bibfnamefont{L.}~\bibnamefont{Chang}},
  \bibinfo{author}{\bibfnamefont{M.}~\bibnamefont{Chen}},
  \bibinfo{author}{\bibfnamefont{X.-q.} \bibnamefont{Li}},
  \bibinfo{author}{\bibfnamefont{Y.-x.} \bibnamefont{Liu}}, \bibnamefont{and}
  \bibinfo{author}{\bibfnamefont{K.}~\bibnamefont{Raya}}, \bibinfo{journal}{Few
  Body Syst.} \textbf{\bibinfo{volume}{62}}, \bibinfo{pages}{4}
  (\bibinfo{year}{2021}), \eprint{1912.08339}.

\bibitem[{\citenamefont{Dominguez et~al.}(1993)\citenamefont{Dominguez,
  Schilcher, and Wu}}]{Dominguez:1993rg}
\bibinfo{author}{\bibfnamefont{C.~A.} \bibnamefont{Dominguez}},
  \bibinfo{author}{\bibfnamefont{K.}~\bibnamefont{Schilcher}},
  \bibnamefont{and} \bibinfo{author}{\bibfnamefont{Y.~L.} \bibnamefont{Wu}},
  \bibinfo{journal}{Phys. Lett. B} \textbf{\bibinfo{volume}{298}},
  \bibinfo{pages}{190} (\bibinfo{year}{1993}).

\bibitem[{\citenamefont{Gershtein et~al.}(1995)\citenamefont{Gershtein,
  Kiselev, Likhoded, and Tkabladze}}]{Gershtein:1994dxw}
\bibinfo{author}{\bibfnamefont{S.~S.} \bibnamefont{Gershtein}},
  \bibinfo{author}{\bibfnamefont{V.~V.} \bibnamefont{Kiselev}},
  \bibinfo{author}{\bibfnamefont{A.~K.} \bibnamefont{Likhoded}},
  \bibnamefont{and} \bibinfo{author}{\bibfnamefont{A.~V.}
  \bibnamefont{Tkabladze}}, \bibinfo{journal}{Phys. Rev. D}
  \textbf{\bibinfo{volume}{51}}, \bibinfo{pages}{3613} (\bibinfo{year}{1995}),
  \eprint{hep-ph/9406339}.

\bibitem[{\citenamefont{Bagan et~al.}(1994)\citenamefont{Bagan, Dosch,
  Gosdzinsky, Narison, and Richard}}]{Bagan:1994dy}
\bibinfo{author}{\bibfnamefont{E.}~\bibnamefont{Bagan}},
  \bibinfo{author}{\bibfnamefont{H.~G.} \bibnamefont{Dosch}},
  \bibinfo{author}{\bibfnamefont{P.}~\bibnamefont{Gosdzinsky}},
  \bibinfo{author}{\bibfnamefont{S.}~\bibnamefont{Narison}}, \bibnamefont{and}
  \bibinfo{author}{\bibfnamefont{J.~M.} \bibnamefont{Richard}},
  \bibinfo{journal}{Z. Phys. C} \textbf{\bibinfo{volume}{64}},
  \bibinfo{pages}{57} (\bibinfo{year}{1994}), \eprint{hep-ph/9403208}.

\bibitem[{\citenamefont{Chen et~al.}(2014)\citenamefont{Chen, Steele, and
  Zhu}}]{Chen:2013eha}
\bibinfo{author}{\bibfnamefont{W.}~\bibnamefont{Chen}},
  \bibinfo{author}{\bibfnamefont{T.~G.} \bibnamefont{Steele}},
  \bibnamefont{and} \bibinfo{author}{\bibfnamefont{S.-L.} \bibnamefont{Zhu}},
  \bibinfo{journal}{J. Phys. G} \textbf{\bibinfo{volume}{41}},
  \bibinfo{pages}{025003} (\bibinfo{year}{2014}), \eprint{1306.3486}.

\bibitem[{\citenamefont{Wang}(2013)}]{Wang:2012kw}
\bibinfo{author}{\bibfnamefont{Z.-G.} \bibnamefont{Wang}},
  \bibinfo{journal}{Eur. Phys. J. A} \textbf{\bibinfo{volume}{49}},
  \bibinfo{pages}{131} (\bibinfo{year}{2013}), \eprint{1203.6252}.

\bibitem[{\citenamefont{Zeng et~al.}(1995)\citenamefont{Zeng, Van~Orden, and
  Roberts}}]{Zeng:1994vj}
\bibinfo{author}{\bibfnamefont{J.}~\bibnamefont{Zeng}},
  \bibinfo{author}{\bibfnamefont{J.~W.} \bibnamefont{Van~Orden}},
  \bibnamefont{and} \bibinfo{author}{\bibfnamefont{W.}~\bibnamefont{Roberts}},
  \bibinfo{journal}{Phys. Rev. D} \textbf{\bibinfo{volume}{52}},
  \bibinfo{pages}{5229} (\bibinfo{year}{1995}), \eprint{hep-ph/9412269}.

\bibitem[{\citenamefont{Chang et~al.}(2020)\citenamefont{Chang, Chen, and
  Liu}}]{Chang:2019eob}
\bibinfo{author}{\bibfnamefont{L.}~\bibnamefont{Chang}},
  \bibinfo{author}{\bibfnamefont{M.}~\bibnamefont{Chen}}, \bibnamefont{and}
  \bibinfo{author}{\bibfnamefont{Y.-x.} \bibnamefont{Liu}},
  \bibinfo{journal}{Phys. Rev. D} \textbf{\bibinfo{volume}{102}},
  \bibinfo{pages}{074010} (\bibinfo{year}{2020}), \eprint{1904.00399}.

\bibitem[{\citenamefont{Chen et~al.}(2020)\citenamefont{Chen, Chang, and
  Liu}}]{Chen:2020ecu}
\bibinfo{author}{\bibfnamefont{M.}~\bibnamefont{Chen}},
  \bibinfo{author}{\bibfnamefont{L.}~\bibnamefont{Chang}}, \bibnamefont{and}
  \bibinfo{author}{\bibfnamefont{Y.-x.} \bibnamefont{Liu}},
  \bibinfo{journal}{Phys. Rev. D} \textbf{\bibinfo{volume}{101}},
  \bibinfo{pages}{056002} (\bibinfo{year}{2020}), \eprint{2001.00161}.

\bibitem[{\citenamefont{Davies et~al.}(1996)\citenamefont{Davies, Hornbostel,
  Lepage, Lidsey, Shigemitsu, and Sloan}}]{Davies:1996gi}
\bibinfo{author}{\bibfnamefont{C.~T.~H.} \bibnamefont{Davies}},
  \bibinfo{author}{\bibfnamefont{K.}~\bibnamefont{Hornbostel}},
  \bibinfo{author}{\bibfnamefont{G.~P.} \bibnamefont{Lepage}},
  \bibinfo{author}{\bibfnamefont{A.~J.} \bibnamefont{Lidsey}},
  \bibinfo{author}{\bibfnamefont{J.}~\bibnamefont{Shigemitsu}},
  \bibnamefont{and} \bibinfo{author}{\bibfnamefont{J.~H.} \bibnamefont{Sloan}},
  \bibinfo{journal}{Phys. Lett. B} \textbf{\bibinfo{volume}{382}},
  \bibinfo{pages}{131} (\bibinfo{year}{1996}), \eprint{hep-lat/9602020}.

\bibitem[{\citenamefont{de~Divitiis et~al.}(2003)\citenamefont{de~Divitiis,
  Guagnelli, Petronzio, Tantalo, and Palombi}}]{deDivitiis:2003iy}
\bibinfo{author}{\bibfnamefont{G.~M.} \bibnamefont{de~Divitiis}},
  \bibinfo{author}{\bibfnamefont{M.}~\bibnamefont{Guagnelli}},
  \bibinfo{author}{\bibfnamefont{R.}~\bibnamefont{Petronzio}},
  \bibinfo{author}{\bibfnamefont{N.}~\bibnamefont{Tantalo}}, \bibnamefont{and}
  \bibinfo{author}{\bibfnamefont{F.}~\bibnamefont{Palombi}},
  \bibinfo{journal}{Nucl. Phys. B} \textbf{\bibinfo{volume}{675}},
  \bibinfo{pages}{309} (\bibinfo{year}{2003}), \eprint{hep-lat/0305018}.

\bibitem[{\citenamefont{Allison et~al.}(2005)\citenamefont{Allison, Davies,
  Gray, Kronfeld, Mackenzie, and Simone}}]{Allison:2004hy}
\bibinfo{author}{\bibfnamefont{I.~F.} \bibnamefont{Allison}},
  \bibinfo{author}{\bibfnamefont{C.~T.~H.} \bibnamefont{Davies}},
  \bibinfo{author}{\bibfnamefont{A.}~\bibnamefont{Gray}},
  \bibinfo{author}{\bibfnamefont{A.~S.} \bibnamefont{Kronfeld}},
  \bibinfo{author}{\bibfnamefont{P.~B.} \bibnamefont{Mackenzie}},
  \bibnamefont{and} \bibinfo{author}{\bibfnamefont{J.~N.} \bibnamefont{Simone}}
  (\bibinfo{collaboration}{HPQCD, FNAL Lattice, UKQCD}),
  \bibinfo{journal}{Nucl. Phys. B Proc. Suppl.} \textbf{\bibinfo{volume}{140}},
  \bibinfo{pages}{440} (\bibinfo{year}{2005}), \eprint{hep-lat/0409090}.

\bibitem[{\citenamefont{Lu et~al.}(2016)\citenamefont{Lu, Anwar, and
  Zou}}]{Lu:2016mbb}
\bibinfo{author}{\bibfnamefont{Y.}~\bibnamefont{Lu}},
  \bibinfo{author}{\bibfnamefont{M.~N.} \bibnamefont{Anwar}}, \bibnamefont{and}
  \bibinfo{author}{\bibfnamefont{B.-S.} \bibnamefont{Zou}},
  \bibinfo{journal}{Phys. Rev. D} \textbf{\bibinfo{volume}{94}},
  \bibinfo{pages}{034021} (\bibinfo{year}{2016}), \eprint{1606.06927}.

\bibitem[{\citenamefont{Liu and Ding}(2012)}]{Liu:2011yp}
\bibinfo{author}{\bibfnamefont{J.-F.} \bibnamefont{Liu}} \bibnamefont{and}
  \bibinfo{author}{\bibfnamefont{G.-J.} \bibnamefont{Ding}},
  \bibinfo{journal}{Eur. Phys. J. C} \textbf{\bibinfo{volume}{72}},
  \bibinfo{pages}{1981} (\bibinfo{year}{2012}), \eprint{1105.0855}.

\bibitem[{\citenamefont{Ferretti et~al.}(2012)\citenamefont{Ferretti, Galata,
  Santopinto, and Vassallo}}]{Ferretti:2012zz}
\bibinfo{author}{\bibfnamefont{J.}~\bibnamefont{Ferretti}},
  \bibinfo{author}{\bibfnamefont{G.}~\bibnamefont{Galata}},
  \bibinfo{author}{\bibfnamefont{E.}~\bibnamefont{Santopinto}},
  \bibnamefont{and} \bibinfo{author}{\bibfnamefont{A.}~\bibnamefont{Vassallo}},
  \bibinfo{journal}{Phys. Rev. C} \textbf{\bibinfo{volume}{86}},
  \bibinfo{pages}{015204} (\bibinfo{year}{2012}).

\bibitem[{\citenamefont{Kalashnikova}(2005)}]{Kalashnikova:2005ui}
\bibinfo{author}{\bibfnamefont{Y.~S.} \bibnamefont{Kalashnikova}},
  \bibinfo{journal}{Phys. Rev. D} \textbf{\bibinfo{volume}{72}},
  \bibinfo{pages}{034010} (\bibinfo{year}{2005}), \eprint{hep-ph/0506270}.

\bibitem[{\citenamefont{Li et~al.}(2009)\citenamefont{Li, Meng, and
  Chao}}]{Li:2009ad}
\bibinfo{author}{\bibfnamefont{B.-Q.} \bibnamefont{Li}},
  \bibinfo{author}{\bibfnamefont{C.}~\bibnamefont{Meng}}, \bibnamefont{and}
  \bibinfo{author}{\bibfnamefont{K.-T.} \bibnamefont{Chao}},
  \bibinfo{journal}{Phys. Rev. D} \textbf{\bibinfo{volume}{80}},
  \bibinfo{pages}{014012} (\bibinfo{year}{2009}), \eprint{0904.4068}.

\bibitem[{\citenamefont{Ferretti et~al.}(2013)\citenamefont{Ferretti, Galat\`a,
  and Santopinto}}]{Ferretti:2013faa}
\bibinfo{author}{\bibfnamefont{J.}~\bibnamefont{Ferretti}},
  \bibinfo{author}{\bibfnamefont{G.}~\bibnamefont{Galat\`a}}, \bibnamefont{and}
  \bibinfo{author}{\bibfnamefont{E.}~\bibnamefont{Santopinto}},
  \bibinfo{journal}{Phys. Rev. C} \textbf{\bibinfo{volume}{88}},
  \bibinfo{pages}{015207} (\bibinfo{year}{2013}), \eprint{1302.6857}.

\bibitem[{\citenamefont{Deng et~al.}(2023)\citenamefont{Deng, Ni, Li, and
  Zhong}}]{Deng:2023mza}
\bibinfo{author}{\bibfnamefont{Q.}~\bibnamefont{Deng}},
  \bibinfo{author}{\bibfnamefont{R.-H.} \bibnamefont{Ni}},
  \bibinfo{author}{\bibfnamefont{Q.}~\bibnamefont{Li}}, \bibnamefont{and}
  \bibinfo{author}{\bibfnamefont{X.-H.} \bibnamefont{Zhong}}
  (\bibinfo{year}{2023}), \eprint{2312.10296}.

\bibitem[{\citenamefont{Ferretti and Santopinto}(2014)}]{Ferretti:2013vua}
\bibinfo{author}{\bibfnamefont{J.}~\bibnamefont{Ferretti}} \bibnamefont{and}
  \bibinfo{author}{\bibfnamefont{E.}~\bibnamefont{Santopinto}},
  \bibinfo{journal}{Phys. Rev. D} \textbf{\bibinfo{volume}{90}},
  \bibinfo{pages}{094022} (\bibinfo{year}{2014}), \eprint{1306.2874}.

\bibitem[{\citenamefont{van Beveren and Rupp}(2004)}]{vanBeveren:2003jv}
\bibinfo{author}{\bibfnamefont{E.}~\bibnamefont{van Beveren}} \bibnamefont{and}
  \bibinfo{author}{\bibfnamefont{G.}~\bibnamefont{Rupp}},
  \bibinfo{journal}{Eur. Phys. J. C} \textbf{\bibinfo{volume}{32}},
  \bibinfo{pages}{493} (\bibinfo{year}{2004}), \eprint{hep-ph/0306051}.

\bibitem[{\citenamefont{van Beveren and Rupp}(2003)}]{vanBeveren:2003kd}
\bibinfo{author}{\bibfnamefont{E.}~\bibnamefont{van Beveren}} \bibnamefont{and}
  \bibinfo{author}{\bibfnamefont{G.}~\bibnamefont{Rupp}},
  \bibinfo{journal}{Phys. Rev. Lett.} \textbf{\bibinfo{volume}{91}},
  \bibinfo{pages}{012003} (\bibinfo{year}{2003}), \eprint{hep-ph/0305035}.

\bibitem[{\citenamefont{Coito et~al.}(2011)\citenamefont{Coito, Rupp, and van
  Beveren}}]{Coito:2011qn}
\bibinfo{author}{\bibfnamefont{S.}~\bibnamefont{Coito}},
  \bibinfo{author}{\bibfnamefont{G.}~\bibnamefont{Rupp}}, \bibnamefont{and}
  \bibinfo{author}{\bibfnamefont{E.}~\bibnamefont{van Beveren}},
  \bibinfo{journal}{Phys. Rev. D} \textbf{\bibinfo{volume}{84}},
  \bibinfo{pages}{094020} (\bibinfo{year}{2011}), \eprint{1106.2760}.

\bibitem[{\citenamefont{Hwang and Kim}(2004)}]{Hwang:2004cd}
\bibinfo{author}{\bibfnamefont{D.~S.} \bibnamefont{Hwang}} \bibnamefont{and}
  \bibinfo{author}{\bibfnamefont{D.-W.} \bibnamefont{Kim}},
  \bibinfo{journal}{Phys. Lett. B} \textbf{\bibinfo{volume}{601}},
  \bibinfo{pages}{137} (\bibinfo{year}{2004}), \eprint{hep-ph/0408154}.

\bibitem[{\citenamefont{Simonov and Tjon}(2004)}]{Simonov:2004ar}
\bibinfo{author}{\bibfnamefont{Y.~A.} \bibnamefont{Simonov}} \bibnamefont{and}
  \bibinfo{author}{\bibfnamefont{J.~A.} \bibnamefont{Tjon}},
  \bibinfo{journal}{Phys. Rev. D} \textbf{\bibinfo{volume}{70}},
  \bibinfo{pages}{114013} (\bibinfo{year}{2004}), \eprint{hep-ph/0409361}.

\bibitem[{\citenamefont{Lee et~al.}(2007)\citenamefont{Lee, Lee, Min, and
  Park}}]{Lee:2004gt}
\bibinfo{author}{\bibfnamefont{I.~W.} \bibnamefont{Lee}},
  \bibinfo{author}{\bibfnamefont{T.}~\bibnamefont{Lee}},
  \bibinfo{author}{\bibfnamefont{D.~P.} \bibnamefont{Min}}, \bibnamefont{and}
  \bibinfo{author}{\bibfnamefont{B.-Y.} \bibnamefont{Park}},
  \bibinfo{journal}{Eur. Phys. J. C} \textbf{\bibinfo{volume}{49}},
  \bibinfo{pages}{737} (\bibinfo{year}{2007}), \eprint{hep-ph/0412210}.

\bibitem[{\citenamefont{Guo et~al.}(2008)\citenamefont{Guo, Krewald, and
  Meissner}}]{Guo:2007up}
\bibinfo{author}{\bibfnamefont{F.-K.} \bibnamefont{Guo}},
  \bibinfo{author}{\bibfnamefont{S.}~\bibnamefont{Krewald}}, \bibnamefont{and}
  \bibinfo{author}{\bibfnamefont{U.-G.} \bibnamefont{Meissner}},
  \bibinfo{journal}{Phys. Lett. B} \textbf{\bibinfo{volume}{665}},
  \bibinfo{pages}{157} (\bibinfo{year}{2008}), \eprint{0712.2953}.

\bibitem[{\citenamefont{Zhou and Xiao}(2011)}]{Zhou:2011sp}
\bibinfo{author}{\bibfnamefont{Z.-Y.} \bibnamefont{Zhou}} \bibnamefont{and}
  \bibinfo{author}{\bibfnamefont{Z.}~\bibnamefont{Xiao}},
  \bibinfo{journal}{Phys. Rev. D} \textbf{\bibinfo{volume}{84}},
  \bibinfo{pages}{034023} (\bibinfo{year}{2011}), \eprint{1105.6025}.

\bibitem[{\citenamefont{Badalian et~al.}(2008)\citenamefont{Badalian, Simonov,
  and Trusov}}]{Badalian:2007yr}
\bibinfo{author}{\bibfnamefont{A.~M.} \bibnamefont{Badalian}},
  \bibinfo{author}{\bibfnamefont{Y.~A.} \bibnamefont{Simonov}},
  \bibnamefont{and} \bibinfo{author}{\bibfnamefont{M.~A.}
  \bibnamefont{Trusov}}, \bibinfo{journal}{Phys. Rev. D}
  \textbf{\bibinfo{volume}{77}}, \bibinfo{pages}{074017}
  (\bibinfo{year}{2008}), \eprint{0712.3943}.

\bibitem[{\citenamefont{Dai et~al.}(2008)\citenamefont{Dai, Li, Zhu, and
  Zuo}}]{Dai:2006uz}
\bibinfo{author}{\bibfnamefont{Y.-B.} \bibnamefont{Dai}},
  \bibinfo{author}{\bibfnamefont{X.-Q.} \bibnamefont{Li}},
  \bibinfo{author}{\bibfnamefont{S.-L.} \bibnamefont{Zhu}}, \bibnamefont{and}
  \bibinfo{author}{\bibfnamefont{Y.-B.} \bibnamefont{Zuo}},
  \bibinfo{journal}{Eur. Phys. J. C} \textbf{\bibinfo{volume}{55}},
  \bibinfo{pages}{249} (\bibinfo{year}{2008}), \eprint{hep-ph/0610327}.

\bibitem[{\citenamefont{Hao et~al.}(2022)\citenamefont{Hao, Lu, and
  Zou}}]{Hao:2022vwt}
\bibinfo{author}{\bibfnamefont{W.}~\bibnamefont{Hao}},
  \bibinfo{author}{\bibfnamefont{Y.}~\bibnamefont{Lu}}, \bibnamefont{and}
  \bibinfo{author}{\bibfnamefont{B.-S.} \bibnamefont{Zou}},
  \bibinfo{journal}{Phys. Rev. D} \textbf{\bibinfo{volume}{106}},
  \bibinfo{pages}{074014} (\bibinfo{year}{2022}), \eprint{2208.10915}.

\bibitem[{\citenamefont{Lakhina and Swanson}(2007)}]{Lakhina:2006fy}
\bibinfo{author}{\bibfnamefont{O.}~\bibnamefont{Lakhina}} \bibnamefont{and}
  \bibinfo{author}{\bibfnamefont{E.~S.} \bibnamefont{Swanson}},
  \bibinfo{journal}{Phys. Lett. B} \textbf{\bibinfo{volume}{650}},
  \bibinfo{pages}{159} (\bibinfo{year}{2007}), \eprint{hep-ph/0608011}.

\bibitem[{\citenamefont{L\"u et~al.}(2016)\citenamefont{L\"u, Pan, Wang, Wang,
  and Li}}]{Lu:2016bbk}
\bibinfo{author}{\bibfnamefont{Q.-F.} \bibnamefont{L\"u}},
  \bibinfo{author}{\bibfnamefont{T.-T.} \bibnamefont{Pan}},
  \bibinfo{author}{\bibfnamefont{Y.-Y.} \bibnamefont{Wang}},
  \bibinfo{author}{\bibfnamefont{E.}~\bibnamefont{Wang}}, \bibnamefont{and}
  \bibinfo{author}{\bibfnamefont{D.-M.} \bibnamefont{Li}},
  \bibinfo{journal}{Phys. Rev. D} \textbf{\bibinfo{volume}{94}},
  \bibinfo{pages}{074012} (\bibinfo{year}{2016}), \eprint{1607.02812}.

\bibitem[{\citenamefont{Li et~al.}(2011)\citenamefont{Li, Ji, and
  Ma}}]{Li:2010vx}
\bibinfo{author}{\bibfnamefont{D.-M.} \bibnamefont{Li}},
  \bibinfo{author}{\bibfnamefont{P.-F.} \bibnamefont{Ji}}, \bibnamefont{and}
  \bibinfo{author}{\bibfnamefont{B.}~\bibnamefont{Ma}}, \bibinfo{journal}{Eur.
  Phys. J. C} \textbf{\bibinfo{volume}{71}}, \bibinfo{pages}{1582}
  (\bibinfo{year}{2011}), \eprint{1011.1548}.

\bibitem[{\citenamefont{Godfrey and Kokoski}(1991)}]{Godfrey:1986wj}
\bibinfo{author}{\bibfnamefont{S.}~\bibnamefont{Godfrey}} \bibnamefont{and}
  \bibinfo{author}{\bibfnamefont{R.}~\bibnamefont{Kokoski}},
  \bibinfo{journal}{Phys. Rev. D} \textbf{\bibinfo{volume}{43}},
  \bibinfo{pages}{1679} (\bibinfo{year}{1991}).

\bibitem[{\citenamefont{Micu}(1969)}]{Micu:1968mk}
\bibinfo{author}{\bibfnamefont{L.}~\bibnamefont{Micu}}, \bibinfo{journal}{Nucl.
  Phys. B} \textbf{\bibinfo{volume}{10}}, \bibinfo{pages}{521}
  (\bibinfo{year}{1969}).

\bibitem[{\citenamefont{Le~Yaouanc et~al.}(1973)\citenamefont{Le~Yaouanc,
  Oliver, Pene, and Raynal}}]{LeYaouanc:1972vsx}
\bibinfo{author}{\bibfnamefont{A.}~\bibnamefont{Le~Yaouanc}},
  \bibinfo{author}{\bibfnamefont{L.}~\bibnamefont{Oliver}},
  \bibinfo{author}{\bibfnamefont{O.}~\bibnamefont{Pene}}, \bibnamefont{and}
  \bibinfo{author}{\bibfnamefont{J.~C.} \bibnamefont{Raynal}},
  \bibinfo{journal}{Phys. Rev. D} \textbf{\bibinfo{volume}{8}},
  \bibinfo{pages}{2223} (\bibinfo{year}{1973}).

\bibitem[{\citenamefont{Le~Yaouanc et~al.}(1974)\citenamefont{Le~Yaouanc,
  Oliver, Pene, and Raynal}}]{LeYaouanc:1973ldf}
\bibinfo{author}{\bibfnamefont{A.}~\bibnamefont{Le~Yaouanc}},
  \bibinfo{author}{\bibfnamefont{L.}~\bibnamefont{Oliver}},
  \bibinfo{author}{\bibfnamefont{O.}~\bibnamefont{Pene}}, \bibnamefont{and}
  \bibinfo{author}{\bibfnamefont{J.~C.} \bibnamefont{Raynal}},
  \bibinfo{journal}{Phys. Rev. D} \textbf{\bibinfo{volume}{9}},
  \bibinfo{pages}{1415} (\bibinfo{year}{1974}).

\bibitem[{\citenamefont{Silvestre-Brac and
  Gignoux}(1991)}]{Silvestre-Brac:1991qqx}
\bibinfo{author}{\bibfnamefont{B.}~\bibnamefont{Silvestre-Brac}}
  \bibnamefont{and} \bibinfo{author}{\bibfnamefont{C.}~\bibnamefont{Gignoux}},
  \bibinfo{journal}{Phys. Rev. D} \textbf{\bibinfo{volume}{43}},
  \bibinfo{pages}{3699} (\bibinfo{year}{1991}).

\bibitem[{\citenamefont{Geiger and Isgur}(1991{\natexlab{a}})}]{Geiger:1991ab}
\bibinfo{author}{\bibfnamefont{P.}~\bibnamefont{Geiger}} \bibnamefont{and}
  \bibinfo{author}{\bibfnamefont{N.}~\bibnamefont{Isgur}},
  \bibinfo{journal}{Phys. Rev. D} \textbf{\bibinfo{volume}{44}},
  \bibinfo{pages}{799} (\bibinfo{year}{1991}{\natexlab{a}}).

\bibitem[{\citenamefont{Geiger and Isgur}(1991{\natexlab{b}})}]{Geiger:1991qe}
\bibinfo{author}{\bibfnamefont{P.}~\bibnamefont{Geiger}} \bibnamefont{and}
  \bibinfo{author}{\bibfnamefont{N.}~\bibnamefont{Isgur}},
  \bibinfo{journal}{Phys. Rev. Lett.} \textbf{\bibinfo{volume}{67}},
  \bibinfo{pages}{1066} (\bibinfo{year}{1991}{\natexlab{b}}).

\bibitem[{\citenamefont{Geiger and Isgur}(1997)}]{Geiger:1996re}
\bibinfo{author}{\bibfnamefont{P.}~\bibnamefont{Geiger}} \bibnamefont{and}
  \bibinfo{author}{\bibfnamefont{N.}~\bibnamefont{Isgur}},
  \bibinfo{journal}{Phys. Rev. D} \textbf{\bibinfo{volume}{55}},
  \bibinfo{pages}{299} (\bibinfo{year}{1997}), \eprint{hep-ph/9610445}.

\bibitem[{\citenamefont{Geng et~al.}(2023)\citenamefont{Geng, Cao, and
  Zhu}}]{Geng:2023ffc}
\bibinfo{author}{\bibfnamefont{Y.}~\bibnamefont{Geng}},
  \bibinfo{author}{\bibfnamefont{M.}~\bibnamefont{Cao}}, \bibnamefont{and}
  \bibinfo{author}{\bibfnamefont{R.}~\bibnamefont{Zhu}} (\bibinfo{year}{2023}),
  \eprint{2310.03425}.

\bibitem[{\citenamefont{Tao et~al.}(2022)\citenamefont{Tao, Zhu, and
  Xiao}}]{Tao:2022qxa}
\bibinfo{author}{\bibfnamefont{W.}~\bibnamefont{Tao}},
  \bibinfo{author}{\bibfnamefont{R.}~\bibnamefont{Zhu}}, \bibnamefont{and}
  \bibinfo{author}{\bibfnamefont{Z.-J.} \bibnamefont{Xiao}},
  \bibinfo{journal}{Phys. Rev. D} \textbf{\bibinfo{volume}{106}},
  \bibinfo{pages}{114037} (\bibinfo{year}{2022}), \eprint{2209.15521}.

\bibitem[{\citenamefont{Wang and Zhu}(2019)}]{Wang:2018duy}
\bibinfo{author}{\bibfnamefont{W.}~\bibnamefont{Wang}} \bibnamefont{and}
  \bibinfo{author}{\bibfnamefont{R.}~\bibnamefont{Zhu}}, \bibinfo{journal}{Int.
  J. Mod. Phys. A} \textbf{\bibinfo{volume}{34}}, \bibinfo{pages}{1950195}
  (\bibinfo{year}{2019}), \eprint{1808.10830}.

\bibitem[{\citenamefont{Zhu}(2018)}]{Zhu:2017lwi}
\bibinfo{author}{\bibfnamefont{R.}~\bibnamefont{Zhu}}, \bibinfo{journal}{Nucl.
  Phys. B} \textbf{\bibinfo{volume}{931}}, \bibinfo{pages}{359}
  (\bibinfo{year}{2018}), \eprint{1710.07011}.

\bibitem[{\citenamefont{Qiao et~al.}(2014)\citenamefont{Qiao, Sun, Yang, and
  Zhu}}]{Qiao:2012hp}
\bibinfo{author}{\bibfnamefont{C.-F.} \bibnamefont{Qiao}},
  \bibinfo{author}{\bibfnamefont{P.}~\bibnamefont{Sun}},
  \bibinfo{author}{\bibfnamefont{D.}~\bibnamefont{Yang}}, \bibnamefont{and}
  \bibinfo{author}{\bibfnamefont{R.-L.} \bibnamefont{Zhu}},
  \bibinfo{journal}{Phys. Rev. D} \textbf{\bibinfo{volume}{89}},
  \bibinfo{pages}{034008} (\bibinfo{year}{2014}), \eprint{1209.5859}.

\end{thebibliography}

\end{document}